
\hoffset=.25in
\vbadness=10000
%
%
\font\bg=cmr10 scaled 1920

\font\bf=cmbx10 scaled 1200

\font\it=cmti10 scaled 1200
\font\sl=cmsl10 scaled 1200

\font\tenrm=cmr10 scaled 1200
\font\sevenrm=cmr9
\font\fiverm=cmr7
\font\teni=cmmi10 scaled 1200
\font\seveni=cmmi9
\font\fivei=cmmi7
\font\tensy=cmsy10 scaled 1200
\font\sevensy=cmsy9
\font\fivesy=cmsy7

\font\tenbf=cmbx10 scaled 1200
\font\sevenbf=cmbx7 scaled 1200
\font\fivebf=cmbx5 scaled 1200
\font\tensl=cmsl10 scaled 1200
\font\tentt=cmtt10 scaled 1200
\font\tenit=cmti10 scaled 1200
\catcode`\@=11
\textfont0=\tenrm \scriptfont0=\sevenrm \scriptscriptfont0=\fiverm
\def\rm{\fam\z@\tenrm}
\textfont1=\teni \scriptfont1=\seveni \scriptscriptfont1=\fivei
\def\mit{\fam\@ne} \def\oldstyle{\fam\@ne\teni}
\textfont2=\tensy \scriptfont2=\sevensy \scriptscriptfont2=\fivesy
\def\cal{\fam\tw@}
\textfont3=\tenex \scriptfont3=\tenex \scriptscriptfont3=\tenex
\newfam\itfam \def\it{\fam\itfam\tenit} 
\textfont\itfam=\tenit
\newfam\slfam \def\sl{\fam\slfam\tensl} 
\textfont\slfam=\tensl
\newfam\bffam \def\bf{\fam\bffam\tenbf} 
\textfont\bffam=\tenbf \scriptfont\bffam=\sevenbf
\scriptscriptfont\bffam=\fivebf
\newfam\ttfam  
\textfont\ttfam=\tentt
\catcode`\@=12
%
%
\rm
\hfuzz=10pt \overfullrule=0pt
\vsize 8.75in
\hsize 6in
\def\doublespace{\baselineskip=28pt}

\parskip 6pt
\def\blankline{\par\vskip \baselineskip}
\doublespace
\parindent=1cm
\raggedbottom
\noindent{\bg Path integration in relativistic quantum mechanics}
\blankline
\noindent{Ian H.~Redmount and Wai-Mo Suen}\par
\noindent{\sl McDonnell Center for the Space Sciences}\par
\noindent{\sl Washington University, Department of Physics}\par
\noindent{\sl St.~Louis, Missouri~~~63130--4899~~~USA}
\blankline
\noindent To be published in {\it International Journal of Modern Physics\/}
{\bf A.}
\blankline
\item{}The simple physics of a free particle reveals important features
of the path-integral formulation of relativistic quantum theories.  The
exact quantum-mechanical propagator is calculated here for a particle
described by the simple relativistic action proportional to its proper time.
This propagator is nonvanishing outside the light cone, implying that
spacelike trajectories must be included in the path integral.  The propagator
matches the WKB approximation to the corresponding configuration-space path
integral far from the light cone; outside the light cone that approximation
consists of the contribution from a single spacelike geodesic.  This
propagator also has the unusual property that its short-time limit does not
coincide with the WKB approximation, making the construction of a concrete
skeletonized version of the path integral more complicated than in
nonrelativistic theory.
\vfill\eject
\noindent{\bf 1.~~Introduction}\par\nobreak
The path-integral formulation of relativistic quantum mechanics gives rise
to problems not found in nonrelativistic theory.  Some of these are similar
to problems which arise in attempting to construct a quantum description of
gravity.  In this work we seek to elucidate some of these problems by
considering the much simpler physics of a single, free, relativistic
particle, a system offering the advantage of exact solubility.

The quantum mechanics of a particle can be completely described by a
propagator, given by a functional integral
$$G(x,t;x_0,t_0)=\int_C{\cal D}x\,e^{iS/\hbar}\ .\eqno(1.1)$$
Here $C$ denotes the class of paths included in the integral, and $S$ is the
classical action associated with each path.  In nonrelativistic theory, $C$
properly includes all paths linking spacetime point~$(x,t)$ to~$(x_0,t_0)$
which move forward in time, in the sense that such a class~$C$ in~(1.1) gives
the same propagator as that in the canonical Hamiltonian quantization.  In a
relativistic theory, with Lorentz-invariant action~$S$, the light-cone
structure of the spacetime comes into play:  Should $C$ include all paths
that move forward in time, or only those inside the light cone, i.e., paths
which are always timelike?  If the latter, then the propagator~(1.1) must
vanish outside the light cone of~$(x_0,t_0)$.  If spacelike paths are
included, this will not be so in general, and $G$ will describe propagation
outside the light cone.

Such propagation will be acausal---backward in time in some Lorentz frames.
Whether such a feature is admissible, or perhaps necessary, is a question
which arises in quantum gravity as well as in particle mechanics.  Indeed,
Teitelboim~[1] has argued that a quantum gravity theory cannot be both
covariant and causal; he suggests~[2] retaining causality, although
Hartle~[3] argues that covariance should be preserved instead.  There too
the problem can be framed as a choice of histories to be included in a
path integral:  It is the choice between the class of spacetime histories
which includes those with negative lapse function, i.e., backward time
displacement, and that which does not~[2,3]. (In fact the action for
general relativity can be converted into a form~[4] like that for a
relativistic particle; general relativity is analogous to parametrized
theories for such particles~[2,3,5].)

Here we evaluate the quantum-mechanical propagator for a free,
relativistic point particle.  We show the significance of spacelike
paths by comparing the exact propagator with the WKB approximation
to the formal, non-Gaussian, configuration-space path integral.

Comparison of the exact and WKB expressions for the propagator in
the short-time-interval limit also yields the appropriate measure
for the path integral.  With this a concrete ``skeletonized'' version
of the formal integral can be constructed.  Here too the relativistic
theory encounters complications not found in the nonrelativistic case.

For simplicity we treat a particle in $1+1$-dimensional flat spacetime;
generalization to higher dimensions is straightforward.  Units with
$\hbar=c=1$ are used henceforth.
\blankline
\noindent{\bf 2.~~Propagator and path integral for a free relativistic
particle}\par\nobreak
A single free particle can be described by the relativistic action
$$S=-\int m\,d\tau=-\int m(1-{\dot x}^2)^{1/2}\,dt\ ,\eqno(2.1)$$
where $m$ is the particle's mass, $\tau$ its proper time, and
$(x,t)$ its coordinates in some Lorentz frame, with $\dot x\equiv dx/dt$.
Hence this model has Lagrangian $L=-m(1-{\dot x}^2)^{1/2}$, momentum
$p=\partial L/\partial\dot x=m\dot x(1-{\dot x}^2)^{-1/2}$, and Hamiltonian
$$H=p\dot x-L=+(p^2+m^2)^{1/2}\ .\eqno(2.2)$$
Being nonpolynomial in~$p$, this corresponds to a nonlocal quantum operator;
it is to be interpreted as acting on any wave function
$$\psi(x,t)=\int dk\,e^{ikx}\,\phi(k,t)\eqno(2.3{\rm a})$$
to give~[6,7]
$$H\psi(x,t)=\int dk\,e^{ikx}\,(k^2+m^2)^{1/2}\phi(k,t)\ .\eqno(2.3{\rm b})$$
The sign of $H$ in Eq.~(2.2) emerges unambiguously from the canonical
formalism, given the action~(2.1).  It implies that the quantum-mechanical
description is entirely in terms of positive-frequency functions.  Such a
description is adequate for a noninteracting particle~[6].  [It is formally
the same as the positive-frequency branch of a Klein-Gordon theory.  In the
Klein-Gordon picture interactions could ``scatter the particle into the
negative-frequency branch.''  Teitelboim~[2], however, has constructed a
theory of {\it interacting\/} relativistic point particles based on an action
of form~(2.1), plus interaction terms.]

The propagator for the wave equation $i\dot\psi=H\psi$ is given by the
integral
$$G(x,t;x_0,t_0)=\int_{-\infty}^{+\infty}{dk\over2\pi}e^{ik\Delta x}\,
e^{-i(k^2+m^2)^{1/2}\Delta t}\ ,\eqno(2.4)$$
with $\Delta x\equiv x-x_0$ and $\Delta t\equiv t-t_0$; clearly this is
a solution of the equation, with ``initial value''~$\delta(\Delta x)$
at~$\Delta t=0$.  The same integral expression can be obtained using the
localized states of Newton and Wigner~[7], i.e., as the projection
of the state localized at~$x_0$ at time~$t_0$ on that localized at~$x$
at time~$t$.  Hartle and Kuchar~[5] derive this same ``Newton-Wigner
propagator'' via a phase-space path integral.

A very simple calculation yields the propagator~$G$ in closed form:
The integral~(2.4) can be evaluated by adding a small negative
imaginary part to~$\Delta t$ for convergence.  The result is
$$G=\lim_{\epsilon\to0^+}{m(i\Delta t+\epsilon)\over\pi
\lambda_\epsilon^{1/2}}\,K_1(m\lambda_\epsilon^{1/2})\ ,\eqno(2.5)$$
with $\lambda_\epsilon\equiv(\Delta x)^2+(i\Delta t+\epsilon)^2$ and
$K_1$ the familiar modified Bessel function.  This propagator contains
the complete description of the behavior of a free particle in this
formulation of relativistic quantum mechanics.

The corresponding configuration-space path integral~(1.1) for the
propagator is, formally,
$$G=\int{\cal D}x\,\exp\left[-im\int_{(x_0,t_0)}^{(x,t)}
(1-{\dot x}^2)^{1/2}\,dt\right]\ .\eqno(2.6)$$
The treatment of spacelike paths in this functional integral is clearly
problematic~[2].  If such paths are to be excluded from the integral by
fiat then $G$ must vanish for all $(x,t)$ outside the light cone
of~$(x_0,t_0)$, i.e., for $|\Delta x|>|\Delta t|$.  But expression~(2.5)
does not do this:  Outside the light cone $m\lambda_\epsilon^{1/2}$ is
(nearly) real, and the Bessel function is nonvanishing---decreasing
exponentially for large argument.  Obviously {\it the path integral for a
relativistic particle must include the contributions of spacelike paths\/}
in general.

Such contributions are manifest in the WKB approximation to the propagator.
In such circumstances as the integral~(2.6) is dominated by paths near the
classical trajectory between $(x_0,t_0)$ and $(x,t)$, it is approximated
by the WKB expression~[8]
$$\int{\cal D}x\,e^{iS}\sim\left({i\over2\pi}{\partial^2S_{\rm cl}\over
\partial x\partial x_0}\right)^{1/2}\,e^{iS_{\rm cl}}\ ,\eqno(2.7)$$
with $S_{\rm cl}$ the action evaluated along that classical path.  Here
that path simply has constant speed~$\dot x=\Delta x/\Delta t$.  Hence the
classical action is $S_{\rm cl}=+im\lambda_\epsilon^{1/2}$ ($\epsilon\to0^+$),
and the resulting approximation is
$$G\sim\lim_{\epsilon\to0^+}\left({m(i\Delta t+\epsilon)^2\over
2\pi\lambda_\epsilon^{3/2}}\right)^{1/2}\,e^{-m\lambda_\epsilon^{1/2}}\ .
\eqno(2.8)$$
(Here the $\epsilon$ term serves to specify the phases of the square roots.)
The exact propagator~(2.5) approaches just this form in the
regime~$m|\lambda_\epsilon^{1/2}|\gg1$, i.e., many Compton wavelengths from
the light cone; in other words, as expected, the WKB approximation to the
path integral~(2.6) is accurate when the magnitude of the classical action
is large compared to~$\hbar$.  For $(x,t)$ well outside the light cone
of~$(x_0,t_0)$, the dominant trajectory giving rise to form~(2.8) of the
propagator is the spacelike geodesic (line) between the two points.

These features of the propagator are illustrated in Fig.~1, which shows
the evolution of a simple Gaussian initial wave function via~$G$.  The
extension of the propagator outside the light cone and the coincidence
of the exact and WKB forms away from the light cone are evident.

Unlike familiar nonrelativistic propagators, the relativistic-particle
propagator does not coincide with the WKB form in the limit~$\Delta t\to0$.
In the regime~$|\Delta x|,|\Delta t|\ll m^{-1}$, the exact propagator~(2.5)
takes the form
$$G({\rm exact})\sim{1\over\pi}\lim_{\epsilon\to0^+}
{i\Delta t+\epsilon\over\lambda_\epsilon}\,\left\{1+
O[m^2|\lambda_\epsilon\ln(m\lambda_\epsilon^{1/2})|]\right\}
\eqno(2.9{\rm a})$$
while the WKB approximation~(2.8) takes the form
$$G({\rm WKB})\sim\left({m\over2\pi}\right)^{1/2}\lim_{\epsilon\to0^+}
{i\Delta t+\epsilon\over\lambda_\epsilon^{3/4}}\,\left[1+
O(m|\lambda_\epsilon^{1/2}|)\right]\ .\eqno(2.9{\rm b})$$
At $\Delta t=0$ the former becomes $\delta(\Delta x)$, as required;
the latter does not.  In this disagreement between the $\Delta t\to0$ and
WKB limits, the relativistic-particle propagator resembles certain
curved-space propagators~[9].

The absence of a factor
$e^{iS_{\rm cl}}=1-m\lambda_\epsilon^{1/2}+O(m^2|\lambda_\epsilon|)$ in
form~(2.9a) might lead one to conclude that a Lagrangian, i.e.,
configuration-space, path integral for the relativistic particle cannot be
constructed, as indicated by Hartle and Kuchar~[5].  This can be done,
however, by including appropriate factors in the path-integral measure.  Thus
a skeletonized version of integral~(2.6) can be given as the composition
$$G(x,t;x_0,t_0)=\lim_{N\to\infty}\int_{-\infty}^{+\infty}dx_1\cdots
\int_{-\infty}^{+\infty}dx_N\,G(x,t;x_N,t_N)\cdots G(x_1,t_1;x_0,t_0)\ ,
\eqno(2.10{\rm a})$$
with the infinitesimal-interval propagators
$$G(x_k,t_k;x_{k-1},t_{k-1})=\lim_{\epsilon\to0^+}\left[{m(i\Delta t+
\epsilon)\over\pi\lambda_\epsilon^{1/2}}\,K_1(m\lambda_\epsilon^{1/2})\,
\exp(m\lambda_\epsilon^{1/2})\right]\,e^{iS_{\rm cl}}\ ,\eqno(2.10{\rm b})$$
where $\Delta t$, $\lambda_\epsilon$, and $S_{\rm cl}$ are taken between
$(x_{k-1},t_{k-1})$ and $(x_k,t_k)$.  This form follows directly from the
exact result~(2.5); since $\Delta x$, hence $\lambda_\epsilon$, need not be
small compared to $m^{-1}$ even when $\Delta t$ is, no expansion of that
result is suitable.  The measure factors, those in square brackets preceding
$e^{iS_{\rm cl}}$ in Eq.~(2.10b), are considerably more complicated than their
nonrelativistic counterparts.  This is to be expected since the kinetic
term in the relativistic action is more complicated than the nonrelativistic
term, which is simply quadratic in the velocity.
\blankline
\noindent{\bf 3.~~Conclusions}\par\nobreak
The free, relativistic point particle provides a simple, exactly
soluble example of some remarkable features of path integrals
for relativistic quantum theories.  The Newton-Wigner~[7] propagator
for such a particle can be expressed in the closed form~(2.5).  Although
the corresponding configuration-space path integral~(2.6) cannot be
evaluated exactly [except in the sense that Eq.~(2.5) {\it is\/} the
evaluation of integral~(2.6)], it can be approximated by the formal
WKB expression~(2.8).  The WKB approximation can be calculated without
knowing the precise form of the path integral measure.  It agrees with the
exact result far from the light cone, which substantiates the suitability of
the path-integral formulation of this theory.   The exact and approximate
forms are seen to differ only near the light cone, i.e., where the classical
action is not large compared to~$\hbar$, as expected.

The propagator is manifestly nonvanishing outside the light cone.
Hence the path integral must include the contributions
of spacelike trajectories; in particular spacelike classical paths give
the dominant contribution in the construction of the WKB approximation
far outside the light cone, where that approximation is valid.

The exact and WKB forms of the propagator show a difference between the
short-time and WKB limits, i.e., the limits $\Delta t\to0$ and $\hbar\to0$,
of the relativistic-particle propagator, unlike its nonrelativistic
counterpart. This discrepancy introduces additional complexity into the
relativistic theory, i.e., into the path-integral measure.  This does not
appear to invalidate the configuration-space path-integral formulation~[5],
though it does diminish its intuitive appeal.

The contribution of spacelike paths to the relativistic-particle path
integral, i.e., the fact that the propagator is nonvanishing outside the
light cone, shows that acausality is a feature of at least this simple
relativistic quantum theory.  Strict causality is recovered as a classical
limit:  The propagator falls off exponentially outside the light cone, on a
scale of the particle's Compton wavelength.  It might be argued that a
first-quantized theory is inappropriate, that quantum field theory is called
for.  Nonetheless the same questions of acausality and the proper domain of
histories in the path integral are known to arise in quantum
gravity~[1--3,10].  The example treated here lends support to the
conjecture~[3] that path integrals in that more complex case should also
include the contributions of acausal histories.
\blankline
\noindent{\bf Acknowledgements}\par\nobreak
We thank C.~Bender, M.~Visser, C.~M.~Will, and K.~Young for useful
discussions in the course of this work.  Financial support was
provided by the U.~S. National Science Foundation via Grants
No.~PHY89--06286 and No.~PHY85--13953.
\blankline
\noindent{\bf References}\par\nobreak
\item{[1]}C.~Teitelboim, Phys.~Rev.~Lett {\bf 50,} 705 (1983).
\item{[2]}C.~Teitelboim, Phys.~Rev.~D {\bf 25,} 3159 (1982);
M.~Henneaux and C.~Teitelboim, Ann.~Phys.~(NY) {\bf 143,} 127 (1982).
\item{[3]}J.~B.~Hartle, Phys.~Rev.~D {\bf 38,} 2985 (1988).
\item{[4]}F.~F.~Baierlein, D.~H.~Sharp, and J.~A.~Wheeler,
Phys.~Rev.~{\bf 126,} 1864 (1962).
\item{[5]}J.~B.~Hartle and K.~U.~Kuchar, Phys.~Rev.~D {\bf 34,}
2323 (1986).
\item{[6]}S.~S.~Schweber, {\it An Introduction to Relativistic Quantum
Field Theory,} (Row and Peterson, Evanston, 1961), pp.~54--62.
\item{[7]}T.~Newton and E.~Wigner, Rev.~Mod.~Phys.~{\bf 21,} 400 (1949).
\item{[8]}L.~S.~Schulman, {\it Techniques and Applications of Path
Integration,} (Wiley, New York, 1981), p.~94.
\item{[9]}Schulman, {\it Techniques and Applications of Path
Integration\/} (Ref.~8), pp.~214--218.
\item{[10]}J.~J.~Halliwell, Phys.~Rev.~D {\bf 38,} 2468 (1988).
\vfill\eject
\centerline{\sl Figure Caption}
\blankline
FIG.~1.  Real (a) and imaginary (b) parts of a wave function propagated via
the relativistic-particle propagator~$G$.  The wave function with initial
form~$\psi(x,0)=\exp(-m^2x^2)$ is shown at time~$t=10m^{-1}$.  The solid
curves are the exact wave function, the dotted that obtained using the
WKB approximation to the propagator.  The values of~$\psi$ are in units
of~$m$.
\vfill\eject\end